\documentstyle[sprocl,psfig]{article}

\def\Journal#1#2#3#4{{#1} {\bf #2}, #3 (#4)}

\def\NPB{{\em Nucl. Phys.} B}

\def\PRL{\em Phys. Rev. Lett.}

\def\be{\begin{equation}}
\def\ee{\end{equation}}
\def\bea{\begin{eqnarray}}
\def\eea{\end{eqnarray}}

\begin{document}

\title{SUMMARY OF EXPERIMENTAL ELECTROWEAK PHYSICS\footnote{
For proceedings of the 17th International Workshop on Weak Interactions and
Neutrinos (WIN99), Cape Town, South Africa, January 24-30, 1999.}}

\author{L. NODULMAN}

\address{HEP Division, Argonne National Laboratory, \\9700 S Cass Ave., 
Argonne, IL 60435, USA\\E-mail: ljn@anl.gov} 


\maketitle\abstracts{ Progress continues on many fronts of experimental
testing of electroweak symmetry breaking.  Updates were presented on
LEP, SLC, Brookhaven g-2 ring, Tevatron Collider,  HERA, CESR and
Tevatron neutrino experiments.
Perhaps most exciting is the Higgs search at LEP2, complementing
the indirect constraints. However,
the standard model with one Higgs doublet remains viable.}

\section{Introduction}

Electroweak physics had its experimental beginning in inelastic 
neutrino scattering neutral current measurements.  
Particular measurements may be interpreted directly or
combined in global fits to constrain the Higgs mass and possibilities for 
new physics.  The increasingly precise magnetic moment measurement of the
muon at Brookhaven will limit nonstandard possibilities.\cite{roberts} 

The $Z$ mass measurement 
has developed a precision in the same league with $G^{\mu}_F$
and $\alpha_{EM}$, and the shape and decays show that we understand the decay
process as well as what states are available.\cite{renton} 
For example there is room
for only three neutrinos. The various $Z$ asymmetries from 
LEP \cite{duckeck} and SLC \cite{rowson} give the strongest indirect constraint
on the Higgs mass. With $e^+e^- \rightarrow hadrons$ measurements at
BES \cite{blondel} and Novosibirsk \cite{steinhauser} complementing or confirming
PQCD calculations,\cite{steinhauser} the precision of $\alpha_{EM}$ evolved to
the $Z$ mass has improved, and this constraint is becoming stronger. 

The combination of $W$ mass \cite{straessner,demarteau} 
and top quark mass \cite{konigsberg} is more of a check at the moment. 
The Tevatron analyses for $W$ and top masses with existing data are
becoming mature, and substantial improvement will come with data from
the next run, which should start in 2000. Considerable improvement
on the $W$ mass is anticipated from the LEP collaborations with
recent data, and more data at higher energy is coming.

New strategies in neutrino scattering make neutral current measurements
interesting,\cite{johnson} and deep inelastic scattering at HERA is
becoming of interest from the electroweak point of view.\cite{brisson}
The absence of the Higgs particle in direct searches \cite{pepe} 
is becoming as significant
an influence on what possibilities remain as the indirect limits. 

Precision
electroweak studies are continuing on many fronts including $\tau$ 
studies,\cite{jessop} and the pending observation of $\nu_{\tau}$
interactions.\cite{rameika} None of these efforts has allowed us to break
out of the standard framework.  I summarize results as presented, but 
update to Moriond 99 numbers.\cite{lepewwg}

\section{BNL 821 Muon g-2}

The study of the magnetic moment of the muon at CERN was precise enough to
demonstrate the presence of hadronic corrections.\cite{cerngm2} The goal
of the ongoing program at Brookhaven is to become precise enough to demonstrate
the electroweak corrections.  More accurate calculations of the hadronic
corrections are helping to make this realistic.\cite{steinhauser}

The experiment is a muon storage ring
consisting of a continuous, finely adjustable, iron dominated
super-conducting magnet.  Mapping and adjusting the field has been an
ongoing program.  The momentum of the muons is adjusted to minimize the
effect of the embedded electrostatic quadrupoles on muon spin precession.
Decays are observed at instrumented windows around the ring,
with high energy decay electrons acting to spin analyze the muons.

A measurement from an initial run with pion decay injection 
approaches the CERN accuracy.\cite{g2prl}  
The numbers are listed in Table \ref{tab3}. 
This result was limited
by low intensity and detector effects, particularly  due to pion injection, 
as well as the field quality.
In two more recent runs, muon injection was established, detectors
improved, and the field much improved.  The available data should produce
a $\pm1$ ppm measurement.  The fringe field of the inflector magnet will
be improved in order for future runs to reach the goal of $~\pm0.3$ ppm for
both signs.

\begin{table}[!bht]
\caption{Muon g-2 measurements ($\times10^{-9}$).
\label{tab3}}
\vspace{0.2cm}
\begin{center}
\footnotesize
\begin{tabular}{|l|c|}
\hline
Measurement & Result \\
\hline
CERN & $1165923.5 \pm8.4$\\
BNL E821 initial run & $1165925 \pm15$\\
\hline
Standard Model Prediction & $1165916.3 \pm0.8$\\
\hline
\end{tabular}
\end{center}
\end{table}

\section{Measurements of the $Z$}

The precision $Z$ line shape has been an adventure story with significant 
implications. The measurement is
\be
m(Z) = 91.1867 \pm0.0021\  {\rm GeV/c}^2
\ee\be
\Gamma(z) = 2.4939 \pm0.0024\ {\rm GeV}.
\ee
The mass is precise enough to rank with
the weak and electromagnetic couplings as precision input. That nothing
seems missing in $Z$ decay 
places serious constraints on new physics possibilities.
Heavy flavor decay rates were a problem but the popular interpretation 
was otherwise ruled out even 
before the deviation went away. 
Agreement continues in the tail of the $Z$ at LEP2.

There is one residual small discrepancy
seen at LEP and less at Moriond by SLC, 
and that is in the asymmetry in b decays.
At 2.6 $\sigma$ or less, depending on input details, the effect is not 
convincing. 

The $Z$ asymmetries from LEP have had some updates in $\tau$ polarization,
and though some further fine tunings are expected, most of the analyses are
pretty much complete.  SLD has gotten a lot more data recently, and the 
preliminary results for the last two years data dominate the measurement.  
The discrepancy between the SLD $A_{LR}$ and the average of the LEP
effective weak mixing continues but has become a lot less jarring,
as seen in Table \ref{tab1}.  Note that the overall average is pulled
up a bit a by the LEP $b$ asymmetry, and down a bit by SLD $A_{LR}$.
A slight residual discrepancy seems historically appropriate.

\begin{table}[!bht]
\caption{Weak mixing measurements $(sin \Theta_{W eff})$ from $Z$ 
forward backward and polarization asymmetries.
\label{tab1}}
\vspace{0.2cm}
\begin{center}
\footnotesize
\begin{tabular}{|l|c|}
\hline
Measurement & Result \\
\hline
LEP lepton fb & $0.23117 \pm0.00054$\\
LEP $\tau$ pol. $A_{\tau}$ & $0.23202 \pm0.00057$\\
LEP $\tau$ pol. $A_e$ & $0.23141 \pm0.00065$\\
LEP $b$ fb & $0.23223 \pm0.00036$\\
LEP $c$ fb & $0.2321 \pm0.0010$\\
LEP jet charge fb & $0.2321 \pm0.0010$\\
SLD pol. $A_{LR}$ & $0.23109 \pm0.00029$\\
\hline
Average ($\chi^2$/DF 7.8/6) & $0.23157 \pm0.00018$\\ 
\hline
\end{tabular}
\end{center}
\end{table}

The impact of the effective weak mixing measurement is improving as
$\alpha_{EM}(m(Z))$ is better determined with PQCD calculations. 
There is discrepancy with old SPEAR hadron rates but new points from
BES agree. Both PQCD and data driven calculations are agreeing and 
improving.\cite{steinhauser}

\section{Measurements of the $W$}

The Tevatron Collider experiments have advanced the program begun at the
CERN $S\bar{p}pS$ collider, including $W$ mass measurements using
leptonic decay transverse mass.  Updates at Moriond had D\O\  adding
plug electrons, and CDF adding the most recent electron sample. Beyond that,
further improvement will come with the next run, expected to start in 2000.

The LEP experiments have threshold $W$ mass measurements, and increasingly
precise direct reconstruction measurements.  Possible QCD systematics in
the four quark mode are being confronted.  The large sample at $\sqrt{s}=189$
GeV should allow a precision of $\pm40-45$ MeV/c$^2$ when fully analyzed;
the ALEPH and L3 analyses included this sample in the Moriond update.
Further data will continue to be collected through 2000. The measurements are
listed in Table \ref{tab2}.

\begin{table}[!bht]
\caption{$W$ mass measurements.
\label{tab2}}
\vspace{0.2cm}
\begin{center}
\footnotesize
\begin{tabular}{|l|c|}
\hline
Measurement & Result \\
\hline
LEP threshold & $80.400 \pm0.221$\\
LEP $qqqq$ & $80.485 \pm0.103$\\
LEP $\ell\nu qq$ & $80.318 \pm0.073$\\
Tevatron $\ell\nu$ & $80.448 \pm0.062$\\
\hline
Direct Average & $80.410 \pm0.044$\\ 
\hline
Indirect fit (LEPEWWG) & $80.364 \pm0.029$\\
\hline
\end{tabular}
\end{center}
\end{table}

Searches for nonstandard $W$ and $Z$ couplings are now dominated by LEP 
measurements, although D\O\  makes a notable contribution. The large new 
data sample at LEP should improve coupling limits by about a factor of three
when fully analyzed.

\section{Measurements of the Top Quark}

The impact of improving the $W$ precision will be limited by the precision
of the top quark mass measurement. The Tevatron data analyses are largely
complete with the two experiments in the different channels consistent,
as can be seen in Table \ref{tab4}.
A couple of years of new data should improve the precision by a factor of at
least two. 

\begin{table}[!bht]
\caption{Top quark mass measurements. When two errors are given,
the first is statistical and the second systematic.
\label{tab4}}
\vspace{0.2cm}
\begin{center}
\footnotesize
\begin{tabular}{|l|c|}
\hline
Measurement & Result \\
\hline
CDF $\ell\nu qqqq$ & $175.9 \pm4.8 \pm5.3$\\
CDF $\ell\nu\ell\nu qq$ & $167.4 \pm10.3 \pm4.8$\\
CDF $qqqqqq$ & $186.0 \pm10.0 \pm5.7$\\
D\O\  $\ell\nu qqqq$ & $173.3 \pm5.6 \pm5.5$ \\
D\O\  $\ell\nu\ell\nu qq$ & $168.4 \pm12.3 \pm3.6$ \\
\hline
Average & $174.3 \pm5.1$\\
\hline
\end{tabular}
\end{center}
\end{table}
Detailed studies of top production and decay, including
limits on nonstandard decays, have begun.  The expected increase in
statistics at the Tevatron once the upgraded collider and detectors get
going should have a salutary effect on these.

\section{Deep Inelastic Scattering}

\begin{figure}[!bht]
\hspace*{.5in}\psfig{figure=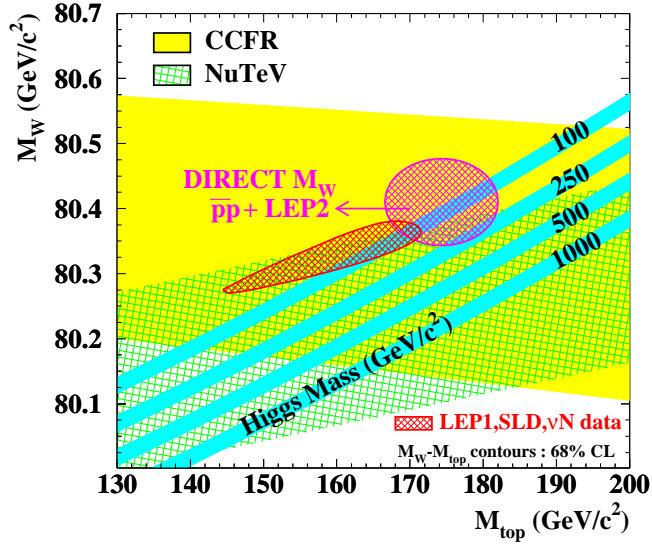,height=3.in}
\caption{$W$ Mass versus Top Mass showing $68\%$ CL allowed regions 
for the combined direct measurements,
the SLC/LEP combined indirect measurement, NuTeV and CCFR with Higgs mass 
predictions.  Note the different slope for the different neutrino analyses.
\label{radish}}
\end{figure}

The NuTeV group has revived the contribution of neutrinos to the 
electroweak program. By using a carefully designed beam, a neutrino beam free
of anti-neutrinos and vice versa allows the difference in neutral to
charged current rates to be used to measure weak mixing. The new data
has similar statistics to CCFR but much improved systematics. The
electroweak physics implications are illustrated in Fig. \ref{radish}.

The HERA experiments still see a small excess at high  $Q^2$, but
not so dramatic as it seemed a year ago.  The data are sufficient to
see propagator effects in NC and CC events. The $t$ channel $W$ mass derived is
compatible with direct measurements, and an electroweak program
is getting started.

A neutrino beam at Fermilab for the NUMI project will have space available
for a near detector. There are some possibilities for PDF studies.\cite{morfin}

\section{The Higgs Search}

The Higgs particle of the minimal standard model has given no direct sign of
its presence. The lower limit on its mass is growing with the LEP energy as
searches are made for the process $e^+e^- \rightarrow Z^* \rightarrow ZH$.
Most signatures involve $H \rightarrow \bar{b}b$, so that $Z$ pairs give an
irreducible background.  Fortunately 
$Z \rightarrow \bar{b}b$ is now well understood. 
The 189 GeV data has now been analyzed by the individual collaborations giving
limits as high as\be
m(H) > 95.2\ {\rm GeV/c}^2 \ 95\%\ {\rm CL}.\ee  The combination of experiments
will give some improvement. With data at 200 GeV, a limit or
discovery reach up to $\sim109$ GeV/c$^2$ is in prospect. 

The recent PDG global fits to electroweak data,\cite{erler} using the
data as of Vancouver 98, gives a most
favored Higgs mass is 107 GeV/c$^2$, so the search is covering quite
interesting territory.  The limit is threatening the viability of 
various popular scenarios for extending the standard model.  

If the Higgs
is still missing at the end of the LEP program, with enough luminosity the
Tevatron detectors could extend the Higgs search. Eventually LHC detectors
will make the search comprehensive.

\section{Tau Physics}

The detailed study of $\tau$ decays involves precise decay parameters, 
neutrino mass limits, and rare decay searches including lepton number violation
and CP violation in the $K^0_s\pi\nu$ angular distribution.
There is plenty of room for non-standard model physics. 
The program being pursued at CLEO will be being joined at Babar and
Belle.

The E872 collaboration at Fermilab is searching for evidence of 
$\nu_{\tau}$ interactions in emulsion at Fermilab.  With part of
the data measured and analyzed, they have six candidates. Although this
corresponds well to expectations, systematic studies to eliminate
background possibilities are pending, but an announcement that 
interactions have been observed is expected soon.

\section{Conclusions}

The simplest scenario for the standard model, with one residual Higgs
particle, remains viable.  In the global fits, the strongest constraint
comes from measurements of $Z$ asymmetries. These dominate the thinness
of the indirect allowed region of Fig. \ref{radish}.
Some updates on these measurements
are pending, but more progress seems likely from $\alpha_{EM}(m(Z))$ 
improvement.

The $W$ mass measurement is improving considerably, and further improvement and
an improved top mass measurement, as will come with the next Tevatron run,
is needed to compete with the $Z$ asymmetries. 

The direct Higgs search is beginning to cut into the indirect fit allowed
region. Perhaps a positive finding will come soon, but it seems like LHC will
be needed to create a contradiction.  Perhaps a contradiction, which would
break us out of our mold, will come from one of the many electroweak studies
which do not directly contribute to the Higgs picture.

\section*{Acknowledgments}

I am grateful to all the speakers whose material I am summarizing,
and to our wonderful hosts. The work was supported in part by
the United States Department of Energy, Division of High Energy
Physics, Contract W-31-109-ENG-38.


\end{document}